\documentclass[useAMS,usenatbib]{mn2e}
\usepackage{psfig}

\newcommand{\plotfull}[1]
           {\centering \leavevmode \psfig{file=#1,width=\textwidth,clip=}}

\def\simlt{\lower.5ex\hbox{$\; \buildrel < \over \sim \;$}}
\def\simgt{\lower.5ex\hbox{$\; \buildrel > \over \sim \;$}}

\title[Non-linear clustering by colour in the 2dFGRS]
{Statistical analysis of galaxy surveys - III: The
  non-linear clustering of red and blue galaxies in the 2dFGRS} 

\author[Croton et al.]{
\parbox[t]{\textwidth}{
Darren J. Croton$^{1}$, 
Peder Norberg$^{2}$,
Enrique Gazta\~{n}aga$^{3}$,
Carlton M. Baugh$^{4}$
}
\vspace*{6pt} \\ 
$^1$Department of Astronomy, University of California, Berkeley, CA, 94720, USA. \\
$^2$SUPA\thanks{The Scottish Universities Physics Alliance}, Institute for Astronomy, University of Edinburgh, Royal Observatory, Blackford Hill, Edinburgh, EH9 3HJ, UK. \\
$^3$Instituto de Ciencias del Espacio (IEEC/CSIC), F. de Ciencies UAB, Torre C5- Par-2a, Bellaterra, 08193 Barcelona, Spain.\\
$^4$Institute for Computational Cosmology, Department of Physics, University of Durham, South Road, Durham DH1 3LE, UK. 
\vspace{-0.5cm}
}

\date{Accepted ---. Received ---;in original form ---}

\begin{document}

\maketitle

\begin{abstract}
We present measurements of the higher-order clustering of red and blue
galaxies as a function of scale and luminosity made from the
two-degree field galaxy redshift survey (2dFGRS).  We use a
counts-in-cells analysis to estimate the volume averaged correlation
functions, $\bar{\xi}_{\rm p}$, as a function of scale up to order
$p=5$, and also the reduced void probability function.  Hierarchical
amplitudes are constructed using the estimates of the correlation
functions: $S_{\rm p} = \bar{\xi}_{\rm p}/\bar{\xi}_{2}^{p-1}$.
We find that:  
1) Red galaxies display stronger clustering than blue galaxies at all
orders measured.
2) Red galaxies show values of $S_p$ that are strongly dependent on
luminosity, whereas blue galaxies show no segregation in $S_p$ within
the errors; this is remarkable given the segregation in the variance.
3) The linear relative bias shows the opposite trend to the
hierarchical amplitudes, with little segregation for the red sequence
and some segregation for the blue.
4) Faint red galaxies deviate significantly from the ``universal''
negative binomial reduced void probabilities followed by all other
galaxy populations.
Our results show that the characteristic colour of a galaxy population
reveals a unique signature in its spatial distribution. Such
signatures will hopefully further elucidate the physics responsible
for shaping the cosmological evolution of galaxies.
\end{abstract}


\section{Introduction}
\label{intro}

The study of the large scale structure of the Universe is now entering
a new phase. The two-degree field galaxy redshift survey (hereafter
2dFGRS; Colless et~al. 2001, 2003) and the Sloan Digital Sky Survey
(SDSS; York et~al. 2000; Adelman-McCarthy et~al. 2006) have yielded
high precision measurements of the power spectrum of galaxy clustering
on large scales (Cole et~al. 2005; Tegmark et~al. 2006; Percival
et~al. 2007; Padmanabhan et~al. 2007). When these measurements are
combined with high angular resolution maps of the temperature
fluctuations in the cosmic microwave background radiation
(e.g. Hinshaw et~al. 2003, 2007), tight constraints can be derived on
many of the parameters in the cold dark matter model (e.g. Sanchez
et~al. 2006; Spergel et~al. 2007). Within this context, the emphasis
in large scale structure studies is shifting to measuring the
clustering signal for samples of galaxies defined by intrinsic
properties such as luminosity, colour, morphology or spectral type,
with the goal of developing our understanding of the physics of galaxy
formation. The SDSS and 2dFGRS catalogues contain sufficiently large
numbers of galaxies over a large enough volume to allow robust
measurements of clustering to be made for such subsamples
(e.g. Norberg et~al. 2001, 2002a; Zehavi et~al. 2002, 2004, 2005;
Madgwick et~al. 2003).

There are many observational clues which point to a dependence of
galaxy properties on their local environment. Dressler (1980) argued
that galaxies follow a morphology-density relation, with the fraction
of early-type galaxies increasing with the local density. Galaxy
clusters have well defined red sequences in the colour-magnitude
relation (Bower, Lucey \& Ellis 1992; Stanford, Eisenhardt \&
Dickinson 1998). Analyses of the 2dFGRS and SDSS data sets have probed
the connection between density and galaxy colour or type over a wider
range of environments than was previously possible to reveal a general
bimodality in the galaxy population (e.g. Lewis et~al.  2002; Hogg
et~al. 2003, 2004; Balogh et~al. 2004).  These results suggest a tight
correlation between the nature of a galaxy and its local environment,
or equivalently the mass of its host dark matter halo. Measures of the
clustering amplitude of different populations of galaxies reveal
different results, indicating that these populations sample the
underlying mass distribution in different ways. Hence, such clustering
measurements can potentially tell us how the efficiency of the galaxy
formation process depends upon halo mass.

The 2dFGRS and SDSS allow us to push measurements of galaxy clustering
beyond the traditional two-point correlation function or power
spectrum. By extending the clustering analysis to higher orders, we
can extract new information about the connection between galaxies and
matter. In previous work, we employed a counts-in-cells analysis to
measure the higher-order correlation functions (Baugh et~al. 2004;
Croton et~al. 2004b) and the void probability function (Croton
et~al. 2004a) for galaxies samples of different luminosity extracted
from the 2dFGRS. We found that the higher-order correlation functions
measured for 2dFGRS galaxies follow a hierarchical scaling pattern,
characteristic of the clustering pattern which results from the growth
of initially Gaussian fluctuations due to gravitational
instability. Croton et~al. (2004a) obtained independent confirmation
of this result when they uncovered an unambiguous universal form for
the void probability function. The values of the correlation functions
do, however, show differences from the expectation for the best
fitting cold dark matter model. Gazta\~{n}aga et~al. (2005) measured
the three-point correlation function without averaging over a cell
volume, using triangles of galaxies with sides of varying length
ratios. They found the first clear evidence for a non-zero second
order or nonlinear bias parameter, suggesting that the relation of
galaxies to the underlying dark matter may be more complicated than
previous analyses had suggested (e.g. Verde et~al. 2002; Lahav
et~al. 2002). Measuring the higher-order correlation functions is a
challenging task, even with surveys of the size of the 2dFGRS and the
SDSS.  Baugh et~al. (2004) found that their measurements of the
correlation function for $L^*$ galaxies in the 2dFGRS were affected on
large scales by the presence of associations of rich clusters; Nichol
et~al. (2006) found similar effects in the SDSS.

In this paper, we extend our earlier work by exploiting the
availability of colour information for the 2dFGRS catalogue. In
Section~2 we briefly describe the 2dFGRS galaxy catalogue, the
counts-in-cells approach we use and the statistics we measure, along
with a reprise of how higher-order clustering measurements can be used
to make deductions about galaxy bias.  In Section~3 we present our
results for the higher-order clustering of 2dFGRS galaxies selected by
both luminosity and colour. We give a simple interpretation of these
results in Section~4 and present a summary in Section~5.

\section{Data and analysis}
\label{method}

The methodology we use is identical to that described by Baugh
et~al. (2004) and Croton et~al. (2004a; 2004b). These authors measured
the higher-order clustering and the void probability function for
galaxies as a function of luminosity, analysing volume limited samples
drawn from the 2dFGRS. In this paper we extend this earlier work to
consider samples defined by galaxy colour in addition to luminosity,
as we did for the case of the 3-point correlation function in
Gazta\~naga et al. (2005).  Full details of the clustering
measurements and a complete discussion of their interpretation can be
found in the above references; in this section, for completeness, we
provide a brief outline of the 2dFGRS (\S~2.1), the statistics
measured (\S~2.2 and \S~2.3) and give a recap of the implications of
the higher-order clustering statistics for galaxy bias (\S~2.4).

\subsection{The 2dFGRS galaxy catalogue}
\label{survey}

Our analysis employs the completed 2dFGRS (Colless et~al. 2001; 2003)
which contains a total of 221,414 unique, high quality galaxy
redshifts down to a nominal magnitude limit of $b_{\rm
J}\!\approx\!19.45$ with a median redshift $z\!\approx\!0.11$. In
addition to $b_{\rm J}$-band magnitudes, $R_{\rm F}$-band images have
now been scanned, allowing a $b_{\rm J}-R_{\rm F}$ colour to be
defined for each galaxy.  To maximise the volume sampled, we restrict
ourselves to regions with spectroscopic completeness in excess of
$50\%$; however, in practice, the typical completeness is much higher
than this in the final 2dFGRS, with the mean completeness in spheres
of radius $<10h^{-1}$Mpc better than $\sim 80\%$.  The remaining
incompleteness is accounted for using the volume weighting corrections
described in Croton et~al. (2004b).

In order to construct a volume limited subsample from a flux limited
redshift survey, it is necessary to model the redshift dependence of a
galaxy's luminosity or magnitude in the passband in which the survey
selection is defined.  We apply the colour dependent $k+e$-correction
model of Cole et~al. (2005) to account for the average change in
galaxy magnitude due to redshifting of the $b_{\rm J}$-filter bandpass
(the ``k-correction'') and also the associated typical galaxy
evolution (the ``e-correction''). The purpose of defining a volume
limited sample is to isolate galaxies of similar intrinsic luminosity
in a sample with a simple radial selection function to facilitate
clustering measurements (see Norberg et~al. 2001; 2002a).  Once such a
sample has been constructed, we can subdivide it by galaxy colour,
using the $b_{\rm J}$ and $R_{\rm F}$-band photometry available for
2dFGRS data. Cole et~al. (2005) analysed the rest frame $b_{\rm
J}-R_{\rm F}$ colour distribution of galaxies in the 2dFGRS and found
a clear division at a rest frame colour of ${\rm b_J}-r_{\rm
F}=1.07$. The bimodality in the colour distribution about this
reference point leads to natural definition of ``red'' and ``blue''
galaxies. This separation by colour is similar, though by no means
identical to the selection by spectral type employed in previous
2dFGRS analyses (Norberg et~al. 2002a; Madgwick et~al. 2002, 2003;
Conway et~al. 2005; Croton et~al. 2005; see figure~2 of Wild
et~al. 2005).

The volume limited samples used in this paper are listed in Table~1
(see also table 1 of Croton et~al. 2004b for further properties of
these samples.  Note that the improved $k+e$-correction used here has
resulted in small differences to the total number of galaxies falling
within each volume limited boundary when compared with our previous
work).  Our analysis covers a wide range of luminosity, from
$\sim$~0.2$\rm L_{b_{\rm J}}^{*}$to $\sim$~4~$\rm L_{b_{\rm J}}^{*}$
(the mean effective luminosities of the volume limited samples differ
by a factor of 6 from the faintest to brightest). The error bars
plotted on our measurements in each of the figures are jackknife
estimates derived by subdividing each sample into 20 areas on the sky.
Our measurements are strongly correlated from bin to bin, so it is
essential to take this into account when fitting models to the
measurements. For the purpose of deriving confidence intervals on
fitted parameters, we compute a full covariance matrix using the
ensemble of mock 2dFGRS catalogues described by Norberg
et~al. (2002b). See section 4 for further details on how errors from
the mocks are used in our analysis.

\subsection{The distribution of counts-in-cells and its moments}

The clustering statistics we employ require an accurate measurement of
the count probability distribution function (CPDF) for the red and
blue galaxy populations in our volume limited samples. This is done
using a counts-in-cells (CiC) analysis. The CPDF for a given smoothing
scale is measured by throwing down a large number ($2.5 \times 10^7$)
of spheres of radius $R$ within the survey volume.  The probability of
finding exactly $N$ galaxies within a sphere of this scale is given
by:
\begin{equation}
P_{\rm N}(R) = \frac{N_{\rm N}}{N_{\rm T}}~.
\label{eq:cpdf}
\end{equation} 
Here $P_{\rm N}(R)$ is the CPDF at the given scale $R$, where $N_{\rm
N}$ is the number of spheres containing $N$ galaxies out of a total of
$N_{\rm T}$ spheres used. From the CPDF, the clustering moments of the
distribution can be calculated directly. For example, the
volume-averaged 2-point correlation function is given by
\begin{equation}
\bar{\xi}_2(R) = \left[\sum_{N=0}^{\infty} P_{\rm N} (R)~ 
  (N/\bar{N}-1)^2 \right] - 1/\bar{N} ,
\end{equation}
where the mean number of galaxies expected in the sphere, $\bar{N}$,
is simply
\begin{equation}
\bar{N}= \sum_{N=0}^{\infty} N P_{\rm N}(R)~.
\end{equation}
Expressions for the higher-order moments in terms of the CPDF are
given in Appendix~A of Gazta\~naga (1994).

In the hierarchical model of galaxy clustering, all higher-order
correlations can be expressed in terms of the 2-point function,
$\bar{\xi}_2$, and dimensionless scaling coefficients, $S_{\rm p}$
(see Bernardeau et~al. 2002):
\begin{equation}
\label{eq:sp}
\bar{\xi}_{\rm \,p} = S_{\rm p} ~\bar{\xi}_2^{\,p-1}.
\end{equation}
Traditionally, $S_3 = \bar{\xi}_3 / \bar{\xi}_2^{\,2}$ is referred to
as the \emph{skewness} of the distribution and $S_4 = \bar{\xi}_4 /
\bar{\xi}_2^{\,3}$ as the \emph{kurtosis}.  Given that both
$\bar{\xi}_2$ and $\bar{\xi}_{p}$ ($p>2$) can be evaluated using the
CPDF, the hierarchical amplitudes $S_{p}$ can readily be measured and
the scaling behaviour of Eq.~\ref{eq:sp} tested (i.e. the validity of
the $p-1$ power law dependence for higher-order correlations; see
Baugh et~al. 2004).

\begin{table*}
  \centering
  \footnotesize
  \caption[]{ The basic properties of the galaxy samples used in this
paper (columns 1-3) and the best fitting relative linear bias $b_{\rm
r}$ (Eq.~\ref{eq:br}; column 4) and relative non-linear bias $c_2'$
(Eq.~\ref{eq:c2'}; column 5).  Also shown are the best fitting $S_p$
measurements to the hierarchical ratios plotted in Fig.~\ref{fig2}
(columns 6-8).  The three sections give the properties and results for
red galaxies, blue galaxies, and all colours combined (included to
facilitate comparison with earlier work). All errors correspond to the
95~\% confidence interval (2-sigma) for fits carried out using
measurements in the range $R=4.5-14h^{-1}$Mpc.}
  \begin{tabular}{ccccccccc} 
    \hline \hline
 population &     \multicolumn{2}{c}{Mag. range} & {$N_{\rm gal}$} & {$b_r$} & {$c_2'$} & {$S_3$} & {$S_4$}& {$S_5$} \\
            &      \multicolumn{2}{c}{\small $M_{b_{\rm J}}$-$5\log_{10}h$} & & (Eq.~\ref{eq:br}) & 
(Eq.~\ref{eq:c2'}) & & & \\
           \hline \\
RED GALAXIES & -18.0 & -19.0 &  7710 & $0.97 \pm 0.06$ & $0.25 \pm 0.15$  & $2.83\pm0.37 $ & $13.5\pm4.5 $ & $80\pm50$ \\
             & -19.0 & -20.0 & 18693 & $0.89 \pm 0.04$ & $0.13 \pm 0.16$  & $2.69\pm0.34 $ & $12.2\pm4.3 $ & $74\pm50$ \\
             & -20.0 & -21.0 & 15147 &         1       &         0        & $2.04\pm0.15 $ & $6.4\pm2.4 $ & $28\pm15$ \\
           \hline \\
BLUE GALAXIES & -18.0 & -19.0 & 14086 & $0.84 \pm 0.05$ & $-0.11 \pm 0.09$   & $2.00\pm 0.13 $ & $5.35\pm1.78 $ & $ 11.0 \pm 6.5 $ \\
              & -19.0 & -20.0 & 22499 & $0.90 \pm 0.04$ & $-0.06 \pm 0.10 $  & $2.09\pm 0.14 $ & $6.66\pm1.47 $ & $ 29\pm 23$ \\
              & -20.0 & -21.0 & 15125 & 1               &     0              & $1.83\pm 0.13 $ & $4.28\pm0.94 $ & $ 10.0\pm 5.5$ \\
           \hline \\
RED+BLUE GALAXIES   & -18.0 & -19.0 & 21796 & $0.85 \pm 0.06$ & $0.02 \pm 0.12$   & $2.44\pm 0.32 $ & $9.46\pm3.15 $ & $ 46 \pm 33 $ \\
              & -19.0 & -20.0 & 41192 & $0.88 \pm 0.04$ & $0.00 \pm 0.11 $  & $2.35\pm 0.30 $ & $8.77\pm3.27 $ & $ 44\pm 36$ \\
              & -20.0 & -21.0 & 30272 & 1               &     0              & $2.01\pm 0.14 $ & $5.97\pm1.31 $ & $ 24\pm 13$ \\
   \hline \hline
  \end{tabular}
\label{table1}
\end{table*}

\subsection{The void probability function}

A complementary way in which to study the hierarchical clustering
paradigm is through the reduced void probability function (VPF). Put
simply, the reduced VPF, $\chi$, is a parametrisation of $P_0$, the
probability of finding an empty sphere in the galaxy distribution, in
terms of the expectation from a purely Poisson galaxy distribution:
\begin{equation}
\label{eq:chi}
\chi\ =\ -\ln(P_0)\ /\ \bar{N}~.
\end{equation}
White (1979) derived Eq.~\ref{eq:chi} after writing the void
probability function as a power series expansion in the moments of the
CPDF ($\bar{\xi}_{\rm p}$) to \emph{all} orders.  Under the
hierarchical ansatz of Eq.~\ref{eq:sp}, this expansion can be
expressed as a function of $\bar{N} \bar{\xi}_{2}$ only:
\begin{equation}
\chi(\bar{N} \bar{\xi_{2}}) = \sum_{p=1}^{\infty} \frac{S_{\rm p}}{p!} (-\bar{N}
\bar{\xi_{2}})^{p-1}~. 
\label{eq:vpf}
\end{equation}
Therefore, if the hierarchical assumption is valid, a plot of $\chi$
as a function of $\bar{N} \bar{\xi}_{2}$ should produce a universal
curve for galaxy catalogues with different mean densities and
clustering properties, assuming common hierarchical amplitudes
for the different populations (we will test this assumption in
Section~\ref{sec:sp}). Note that clustered tracers which deviate
strongly from a Poisson distribution will have VPF values $\chi < 1$.

The precise form of the reduced VPF is set by the hierarchical
amplitudes, $S_{\rm p}$, for which different models of clustering
predict different values.  Here we summarise only the most successful
model, the \emph{negative binomial model}, which does very well at
reproducing the measurements from the 2dFGRS for both the VPF (Fig.~3
in Croton et~al. 2004b) and the hierarchical amplitudes, $S_{\rm p}$
(Table~1 in Baugh et al. 2004).  The reduced VPF and hierarchical
amplitudes for the negative binomial model are
\begin{eqnarray}
\chi &=& \ln (1+\bar{N} \bar{\xi}_{2})/\bar{N} \bar{\xi}_{2} \\ \nonumber
S_{\rm p} &=& (p-1)!  
\end{eqnarray}
Further analytic models, like the minimal and thermodynamic, are
explored in Croton et~al. (2004b).

\subsection{Higher-order moments and relative bias}
\label{results}

The distribution of galaxies could be quite different from that of the
underlying dark matter. One could conceive of physical processes which
could lead to a dependence of the efficiency of galaxy formation on
the mass and perhaps the environment of dark matter hales
(e.g. Croton, Gao \& White 2007). If this is indeed the case, galaxies
should be regarded as {\em biased} tracers of the matter distribution.

A simple model for galaxy bias was introduced by Fry \& Gazta\~{n}aga
(1993).  These authors assumed that the density contrast in the galaxy
distribution, $\delta^{\rm G}$, can be expressed as a general
non-linear function of the density contrast of the dark matter,
$\delta^{\rm DM}$, so that: $\delta^{\rm G} = F[\delta^{\rm DM}]$. For
density fluctuations smoothed on large enough scales so that the
matter density contrast is of the order unity or smaller, this
relation can be expanded in a Taylor series:
\begin{equation}
\delta^{\rm G}  = \sum_{k=0}^{\infty} \frac{b_{k}}{k!} (\delta^{\rm DM})^{k}\,.
\end{equation}
On scales where the variance, $\bar{\xi}^{\rm DM}$, is small, the
leading order contribution to the variance is dominated by the
linear term:
\begin{equation}
\bar{\xi}_{2}^{\rm G} = b_{1}^{2} ~\bar{\xi}_{2}^{\rm DM} 
+ {\sc O}[\bar{\xi}_{2}^{\rm DM}]^2, 
\label{eq:b1}
\end{equation} 
where $b_1$ is the so-called ``linear bias'', $b$.  The leading order
term in the expansion for the skewness, $S_3$, is:
\begin{equation}
S^{\rm G}_{3} = \frac{1}{b_{1}} \left(S^{\rm DM}_{3} + 3 c_{2}\right)\,, 
\label{eqnfg93}
\end{equation} 
where we use the notation $c_{2} = b_{2}/b_{1}$ (expressions for the
hierarchical amplitudes up to $p=7$ are given in Fry \& Gazta\~{n}aga
1993). $S^{\rm DM}_{3}$ encodes the non-linear gravitational evolution
of the matter distribution from Gaussian initial conditions.  Hence,
two non-linear effects are present in $S^{\rm G}_{3}$: gravity
(i.e. as contained in $S^{\rm DM}_{3}$) and galaxy bias (i.e. as
quantified by $c_2$). Both terms are modulated by the linear bias,
$b_1$. Therefore, in general, the interpretation of the skewness
measured in the galaxy distribution is not trivial.  However, on
weakly non-linear scales we have detailed models of what to expect for
the skewness of the dark matter (see, for example, Bernardeau
et~al. 2002; calculations have also been carried out for non-standard
gravitational models and cosmologies -- see Gazta\~{n}aga \& Lobo
2001).  We can therefore hope to learn about non-linear galaxy biasing
if we can measure the skewness in the galaxy distribution, $S^{\rm
G}_{3}$, on scales on which the underlying fluctuations are only
weakly nonlinear.

It is useful to define a {\emph{relative}} bias to aid with the
interpretation of measurements of higher-order clustering.  The
relative bias describes the change in clustering signal compared to
that measured for a reference sample.  Using Eq.~\ref{eq:b1} as a
guide, we define the relative linear bias of a sample, $b_{\rm
r}=b_1/b^*_1$, as the square root of the ratio of the $2$-point
correlation function (or variance) measured for the sample, relative
to that found for the reference sample, denoted by an asterisk:
\begin{equation}
b_{\rm r} \equiv {b_1\over{b^*_1}} 
= \Big( \frac{\bar{\xi}_{2}^{\,\rm G}}{\bar{\xi}_{2}^{\,\rm G *}} \Big) ^{1/2}\,.
\label{eq:br}
\end{equation}
Our reference sample is the volume limited population of galaxies in
the magnitude range $-20> M_{b_{\rm J}}-5\log_{10}h > -21$, which is
one magnitude brighter than the one used in Croton et~al. (2004a). The
reason for choosing a brighter sample is directly related to the
influence the large coherent superstructures have on the $L^*$ sample:
it makes more sense to use as reference a sample which is not
systematically affected by the presence of such rare massive
structure.  For completeness, in Table~1 we show the relative bias
results for the volume limited ``all colour'' catalogues analysed in
Croton et al. (2004a), however now presented relative to this brighter
reference sample.

On scales for which a linear bias is a good approximation, i.e. when
$b_k \simeq 0$ for $k>1$, we can relate the $S_{\rm p}^{\rm G}$
measured for different galaxy samples regardless of the value of
$S^{\rm DM}_{\rm p}$ for the underlying mass:
\begin{equation} 
S^{\rm G}_{\rm p} = \frac{S^{\rm G *}_{\rm p}}{b^{p-2}_{\rm r}}.
\label{eq:SJlinear}
\end{equation} 
More generally, if galaxy bias is non-linear one can introduce a
measure to quantify the relative non-linear bias:
\begin{equation} 
c_{2}' = \frac{(c_{2}- c^{*}_{2})}{b_1^*} = \frac{1}{3} \left(b_{\rm r} 
S^{\rm G}_{3} - 
S^{\rm G *}_{3} \right), 
\label{eq:c2'}
\end{equation} 
where an asterisk denotes a quantity measured for the reference
sample. In general, if the reference sample is un-biased
(i.e. $b_1^*=1$ and $c_{\rm p}^*=0$), we then have $c_{2}' =c_{2}$ and
$b_r = b_1$ for all samples.

\section{Results}

In this section we present the main results of the paper, which are
inferred from our measurement of the count probability distribution
function (CPDF), as outlined in Section 2, for the samples of red and
blue galaxies listed in Table~1. We present the volume averaged
correlation functions, $\bar{\xi}_{\rm p}$, in \S~3.1, the
hierarchical amplitudes, $S_{\rm p}$, in \S~3.2, and the reduced void
probability function (VPF) in \S~3.3.

\begin{figure*}
\plotfull{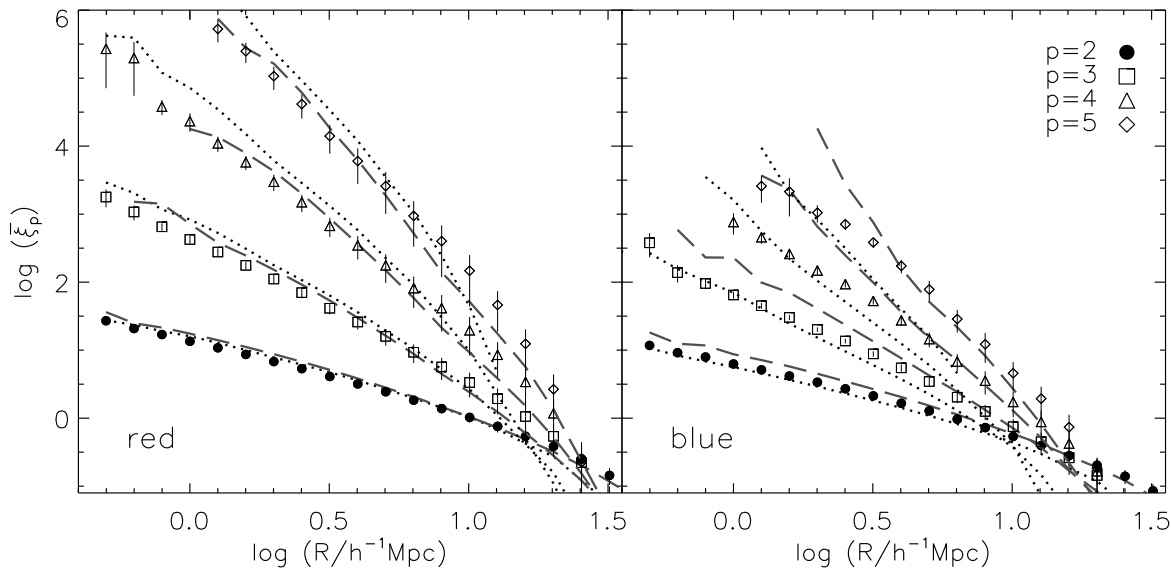}
\caption{
The p-point volume averaged correlation functions as a function of
scale.  The left and right panels show results for red and blue
galaxies separately.  In each panel, the symbols denote galaxies in
the magnitude range $-19>M_{b_{\rm _J}}-5\log_{10} h>-20$, while the
adjacent dashed and dotted lines show the clustering measured for
galaxies defined by magnitude ranges which are, respectively, one
magnitude brighter and fainter (Table~\ref{table1}). Different symbols
show different orders of clustering, as indicated by the key.}
\label{fig1}
\end{figure*}

\subsection{The higher-order correlation functions} 
\label{sec:xip}

Fig.~\ref{fig1} shows the volume averaged correlation functions for
orders $p=2$--$5$ as a function of smoothing scale. The left hand
panel shows the measurements for red galaxies and the right panel for
blue galaxies.  In both panels the symbols show the results for
samples in the magnitude range $-19>M_{b_{\rm J}}-5\log_{10} h>-20$
(i.e. approximately $L_{b_{\rm J}}^{*}$), while the adjacent dashed
and dotted lines show samples which are one magnitude brighter and
fainter respectively.  The order of volume averaged correlation
function is indicated by the symbol type shown in the legend.

Red galaxies clearly show a larger clustering amplitude than the
equivalent population of blue galaxies, as expected from earlier
results obtained for the two-point correlation function as a function
of colour (e.g. Zehavi et~al. 2005 in the SDSS) or spectral type
(Norberg et~al.  2002a; Madgwick et~al. 2003 in the 2dFGRS). The
dependence of clustering strength on galaxy luminosity appears to be
the strongest for the blue population, with the faintest blue galaxies
being more weakly clustered than the brightest. The situation is more
complicated for red galaxies; both the faintest and brightest samples
of red galaxies appear to be more strongly clustered than the red
$L^{*}$ sample.  There is even a suggestion that the faintest red
galaxies become the more strongly clustered with increasing order,
although the errors on the measurements are also increasing at a fixed
scale for higher values of $p$. This type of behaviour was noted
previously for the two-point correlation function. Norberg
et~al. (2002b) found an increase in the correlation length of early
spectral types for samples faintwards of $L^*$, with the faintest
early types displaying a similar correlation length to the brightest
early types. Zehavi et~al. (2005) found a similar increase in
clustering strength for faint red galaxies. This trend can be readily
explained; the satellite population in massive clusters is
predominantly made up of faint, red galaxies and clusters are known to
be strongly biased tracers of the mass distribution (e.g. Padilla
et~al. 2004).

\subsection{The hierarchical amplitudes}
\label{sec:sp}

\begin{figure*}
\plotfull{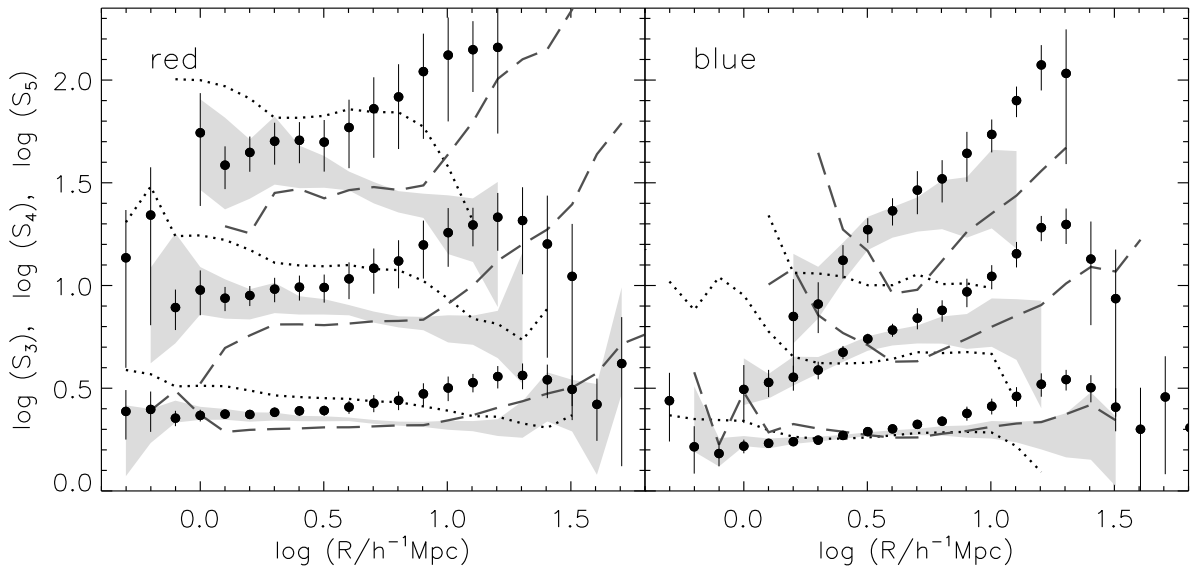}
\caption{
The hierarchical amplitudes, $S_{\rm p}$, plotted as a function of
scale. The left panel shows the results for red galaxies and the right
panel those obtained for blue galaxies. The symbols and lines have the
same meaning as those used in Fig.~\ref{fig1}. The grey shaded regions
show how the results for the $L_{\rm b_J}^{*}$ galaxy sample change
when the two largest superstructures are excluded from the CiC
analysis, as described in the text.  }
\label{fig2}
\end{figure*}

Fig.~\ref{fig2} extends the results presented in the previous
subsection by plotting the hierarchical amplitudes (as defined by
Eq.~\ref{eq:sp}), $S_3$ (the skewness), $S_4$ (the kurtosis) and
$S_5$, as a function of scale. This is done for red galaxies in the
left panel and blue galaxies in the right panel. Again, the results
are shown for samples with different luminosities, with symbols
representing the measurement of galaxies in the magnitude range
$-19>M_{b_{\rm J}}-5\log_{10} h>-20$, and dashed and dotted lines
showing the $S_{\rm p}$ for samples one magnitude brighter and fainter
than this respectively. Red galaxies show a strong segregation of the
hierarchical amplitudes with luminosity at a given clustering order;
the $S_{\rm p}$ for blue galaxies show little dependence on
luminosity.

On large scales, N-body simulations and perturbation theory suggest
that the hierarchical amplitudes for the mass should only vary slowly
with scale in the case of initially Gaussian fluctuations which grow
through gravitational instability (e.g. Juskiewicz, Bouchet \& Colombi
1993; Baugh et~al. 1995; for a summary of the theoretical predictions,
see Bernardeau et~al. 2002).  Observationally, the hierarchical
amplitudes for galaxies have traditionally been measured to be either
approximately constant or decreasing slowly out to progressively
larger smoothing scales (e.g.  Gazta\~{n}aga 1992; Bouchet
et~al. 1993; Gazta\~{n}aga 1994; Hoyle, Szapudi \& Baugh 2000; Frith,
Outram \& Shanks 2006; see Bernardeau et~al. 2002 for a review of
previous measurements).  However, recent work has shown that the
presence of large, rare superstructures in the galaxy distribution can
produce a significant upturn in the value of $S_{\rm p}$ on large
scales (e.g. Szapudi \& Gazta\~{n}aga 1998; Baugh et~al. 2004; Croton
et~al. 2004b; Nichol et~al. 2006).  According to figure 10 from
Gazta\~{n}aga et~al. (2005), the upturn in the hierarchical amplitudes
appears to be equally important when using measurements of the 3-point
correlations function of red and blue galaxies from the 2dFGRS
$-19>M_{b_{\rm J}}>-20$ sample.  Nevertheless, it should be emphasised
that, in the case of the 2dFGRS, the impact of the superstructures on
the measured hierarchical amplitudes is most pronounced in the
$-19>M_{b_{\rm J}}-5\log_{10}h>-20$ volume limited sample and is
essentially negligible in the brighter and fainter samples. This is
due to the superstructures falling entirely within the redshift
interval spanned by the $L^*$ sample. In the case of the fainter
volume limited sample, the superstructures lie mostly beyond the
maximum redshift which defines the sample, whilst in the case of the
brighter sample, the volume covered is much larger and the
significance of the superstructures is correspondingly lower. See
Croton et~al. (2004b), Baugh et~al. (2004) and Gazta\~naga
et~al. (2005) for detailed investigations into how these two
superstructures affect the measurement of higher-order statistics. The
flatness of the earlier results for the $S_{\rm p}$ can be understood
by noting that a) most smaller surveys never intersected with these
rare superstructures of galaxies, b) their influence for larger
surveys, such as the original 2dFGRS parent catalogue, the APM Survey
(i.e. Gazta\~{n}aga 1994) or 2MASS (Frith, Outram \& Shanks 2006), was
diluted by the significant increase in sample volume, and c) typically
any contribution from large coherent superstructures is downgraded in
a flux limited catalogue rather than a volume limited sample due to
the varying selection function.

In Fig.~\ref{fig2}, the grey shaded regions show the results for the
hierarchical amplitudes measured from the $L_{b_{\rm J}}^{*}$ sample
after removing the two large superstructures identified in Baugh
et~al. (2004). The shaded regions indicate the extent of the
$1\!-\!\sigma$ errors on the measurements and should be compared with
the symbols and error bars, which show the results for the same volume
limited sample but {\em without} removing the superstructures from the
CiC analysis. By omitting the two superstructures, the theoretically
expected constant behaviour of $S_{3}$ is restored for larger
smoothing lengths. However, whilst the results change on large scales
for $S_4$ and $S_5$ on removing the superstructures, there is still
significant scale dependence of the $S_{\rm p}$ with opposite
gradients found for red and blue galaxies. From Fig.~\ref{fig2}, the
impression is that the red population is the most strongly affected by
the presence of the superstructures\footnote{A robust quantitative
statement would require the use of more realistic mocks, for which
colour dependent absolute errors on $S_p$ can be derived.}.  This
behaviour should be contrasted with the measurements by Croton et~al.
(2004b) and Baugh et~al. (2004) made using the full L$_{\rm b_J}^{*}$
sample.  In these papers, removing the superstructures from the CiC
analysis, led, to within the accuracy of the measurements, to all the
hierarchical moments up to 5$^{\rm th}$ order becoming roughly
independent of scale.  Our results for red and blue galaxies imply
that the $L^{*}$ results for all galaxies were due to a fortuitous
cancellation of the trends with scale seen for red and blue galaxies.

Finally, it is worth noting that the upturn seen on small scales in
both $S_4$ and $S_5$ for the bright and faint populations of blue
galaxies is not statistically significant: in both cases, the
jackknife errors, if plotted, are fully consistent with no upturn at
all.

\subsection{The reduced void probability function}
\label{sec:vpf}

\begin{figure*}
\plotfull{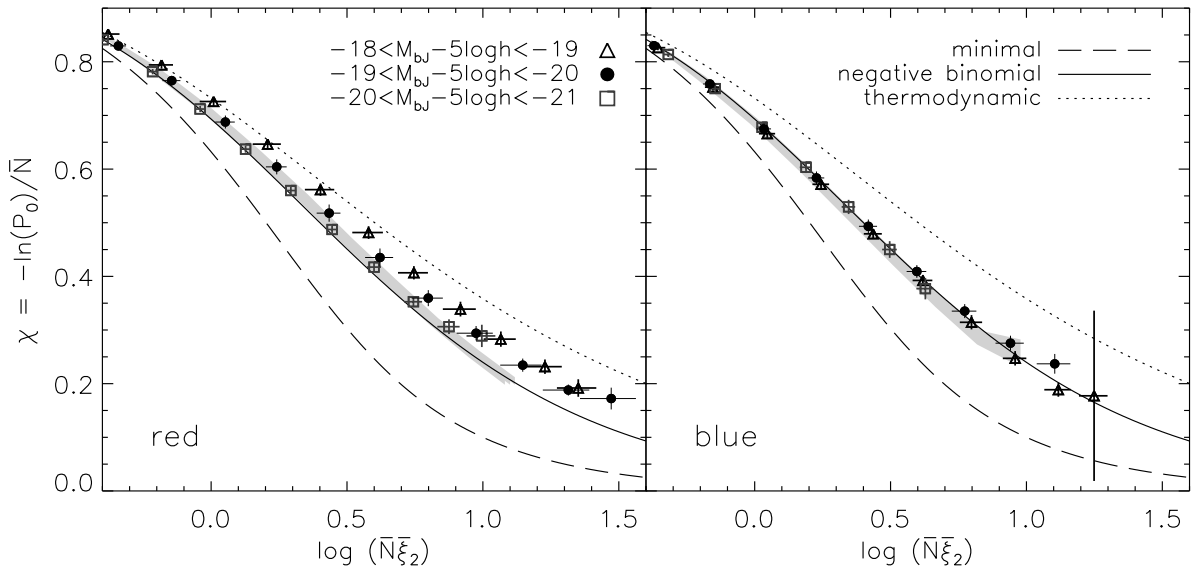}
\caption{
The reduced void probability function plotted as a function of the
scaling variable $\bar{N} \bar{\xi}$.  Left and right panels show the
measurements for the red and blue populations respectively, while the
symbols denote galaxy samples of different luminosity as indicated by
the legend. The grey shaded region shows how the results change for
the $L_{b_{\rm J}}^{*}$ sample when the two largest superstructures
are removed from the analysis.
}
\label{fig3}
\end{figure*}

In Fig.~\ref{fig3} we plot the reduced void probability function (VPF)
for the three luminosity samples (using different symbols as shown by
the legend) and for red and blue galaxy populations (left and right
hand panels respectively).  Previous measurements have revealed that
galaxy samples defined by luminosity (or equivalently number density)
display a universal form for the VPF (e.g. Croton et~al. 2004a; Hoyle
\& Vogeley 2004; Patiri et~al. 2006). This universality also holds for
samples at different redshifts (Conroy et~al. 2005; 2007).  The
recovery of a universal form for the VPF is evidence for the
hierarchical scaling of the higher-order correlation functions of
galaxies (e.g. Eq~\ref{eq:sp}). The actual form of the VPF curve is
well described, in redshift space, by the negative binomial model (see
also Fry 1986 and Gazta\~naga \& Yokoyama 1993).

Here we report a significant departure from the universal scaling of
the VPF. As can be seen in Fig.~\ref{fig3}, the VPF measured for
\emph{red} galaxies deviates from the universal scaling in a way such
that the deviation is strongest for the faintest galaxies considered.
The VPF measured for blue galaxies, on the other hand, shows no
dependence on luminosity, and matches almost perfectly the negative
binomial model across the entire range plotted.  Note that the
deviation seen for the red $L_{b_{\rm J}}^{*}$ population appears to
arise from the contribution of the two superstructures discussed in
the previous subsection. Once these large structures are removed from
the analysis and the VPF is recalculated, the results for the
$L_{b_{\rm J}}^{*}$ sample shift back to the negative binomial model,
as shown by the grey shaded region. On the other hand, the fainter red
galaxy sample (shown by the triangles in Fig.~\ref{fig3}), which is
not affected by the two superstructures, does not agree with the
negative binomial model.  This disagreement persists if we consider
even fainter samples, which we have not plotted in the figure for
clarity.

Our results show that once we have accounted for the difference in
mean density and clustering, faint red galaxies have a
characteristically larger VPF than is measured for the other galaxy
samples.  From Eq.~\ref{eq:vpf}, we expect that samples with a larger
VPF should also also have larger values for the $S_{\rm p}$, and this
is borne out by the results given in Table~1, which are discussed in
the next section.  A high value for $S_{\rm p}$ indicates non-linear
biasing. If one were to simply increase the {\em linear} bias, this
would increase the amplitude of the correlation functions,
$\bar{\xi}_{\rm p}$, but would reduce the hierarchical amplitudes (see
Eq.~\ref{eq:sp} and Eq.~\ref{eq:SJlinear}).  A non-zero non-linear
bias is characteristic of the high density regions of the dark matter
distribution. This is in line with the suggestion made above that the
red faint galaxies could be made up of a significant population of
satellites in clusters.

The reasons behind the success of the negative binomial model for the
VPF are unclear. Vogeley et~al. (1994) showed that the VPF measured in
their simulations only agreed with the negative binomial model when
the clustering pattern was distorted by peculiar motions; the VPF
measured in real space did not match the negative binomial model.
However, our results imply that the peculiar motions of galaxies are
not the primary agent responsible for the success of the negative
binomial model. We find that the faint red galaxies, which are
primarily satellites within galaxy clusters and hence likely to
display large peculiar motions, in fact show the strongest departure
from the VPF predicted by the negative binomial model, rather than
being driven towards this model.

\section{Discussion} 

In this section we give some simple interpretations of the
measurements of the hierarchical amplitudes reported in
Section~\ref{sec:sp}. This requires us to perform fits to the
measurements, in the first case by extracting best fitting values for
the hierarchical amplitudes, $S_{\rm p}$ (as defined in
Eq.~\ref{eq:sp}), and then to constrain the linear and non-linear bias
parameters (see Eq.~\ref{eq:br} and Eq.~\ref{eq:c2'}).

Measures of the hierarchical amplitudes on different smoothing scales
are correlated, a fact which has often been ignored in previous
analyses reported in the literature. The existence of a bin-to-bin
correlation is clearly demonstrated by the response of our estimates
of the hierarchical amplitudes to the removal of the two largest
superstructures in the 2dFGRS. Given this correlation, it is necessary
to construct a covariance matrix for the measurements in order to
carry out meaningful fits.  One way to do this is to use a jackknife
approach in which the survey volume is split into a number of equal
sized subvolumes, on the order of 20 (if not more to construct a
stable covariance matrix). Unfortunately, this is infeasible for our
fainter volume limited samples, as the volume covered in these cases
is too small to be subdivided an appropriate number of times.
Instead, we use the ensemble of mock 2dFGRS catalogues whose
construction is described in Norberg et~al. (2002b). Particles were
chosen from an N-body simulation to represent galaxies using a simple
parametric function of the local smoothed density of dark matter (see
Cole et~al. 1998 for a description of the bias algorithm). The
parameters in this empirical biasing prescription were constrained to
reproduce the typical clustering measured in the full flux limited
2dFGRS (Hawkins et~al. 2003).  Consequently, one limitation of the
mock catalogues is that the luminosity and colour assigned to a
``galaxy'' is independent of the local density of the particle. This
means that the mocks do not display any dependence of clustering on
luminosity or colour. Also, no information about the higher-order
clustering of mock galaxies was used to select the parameters in the
bias prescription. Hence, although the two point correlation function
for mock galaxies agrees very closely with that measured for the full
2dFGRS, there is no guarantee that the higher-order clustering in the
mocks will look like that measured in the real survey.

As we found in our earlier work using the mocks (Croton et~al. 2004b;
Gazta\~{n}aga et~al. 2005), the covariance matrices estimated from the
mocks are stable and give robust estimates of best fitting quantities
and the associated errors. A principal component decomposition of the
covariance matrix reveals that the first few eigenvectors are
typically responsible for the bulk of the variance or signal. We rank
the eigenvectors in order of decreasing variance and retain sufficient
eigenvectors to account for, at least, 95\% of the variance.

We first perform fits to extract values for the scale {\em
independent} hierarchical amplitudes.  Although our results may appear
in some cases to show a strong dependence on cell radius
(Fig.~\ref{fig2}), it is important to bear in mind that we can only
show the diagonal part of the covariance matrix in plots and our
fitting procedure takes into account the covariance between
bins. Moreover, since there is a tradition of fitting constants to
estimates of the $S_{\rm p}$ over a range of scales, it is useful to
repeat such an analysis for our measurements in order to facilitate
comparisons with previous work. We restrict our attention to the
$S_{\rm p}$ values obtained for spheres with radii in the range
$R=4.5-14h^{-1}$Mpc. These are slightly larger scales than we used in
Croton et~al. (2004b), since we aim to avoid scales which may be
affected by shot noise, and our samples, as a result of being split by
colour, typically have half the number of galaxies in each luminosity
bin than before. To help with a comparison to our earlier work we also
re-analyse the ``all colour'' galaxy catalogues across this new
fitting range (note that these galaxies also differ in their improved
$k+e$ corrections, as described in Section~\ref{survey}).  The results
of a one parameter fit ($S_{\rm p}$) to the hierarchical amplitudes
are quoted as a function of luminosity for all galaxy colours and also
the red and blue populations separately in Table~\ref{table1}, in
which we also give the 95~\% confidence interval around the best
fitting value.

Next we constrain the relative bias parameters defined in
Eq.~\ref{eq:br} and Eq.~\ref{eq:c2'}. Recall that in this paper we
define our reference sample to be volume limited galaxies in the
magnitude range $-20> M_{b_{\rm J}}-5\log_{10}h > -21$.  The error in
the relative linear bias is straightforward to find using the mocks
and the results are given in Table~1. Obtaining a robust fit for the
nonlinear bias, $c_{2}^{'}$, is more subtle. By construction, the
mocks will return a zero nonlinear bias\footnote{Without luminosity
segregation, the expectation value for the mocks is $c_{2}^{'} \approx
0$.}, whereas the expectation value for 2dFGRS samples can be
anything.  Therefore, in order to get a reliable estimate of the error
on the best fitting nonlinear bias using the mocks, we need to take an
indirect approach. We fit instead the combination $b_{\rm r} S^{\rm
G}_{3} $, for which the mock catalogues are in rather good agreement
with the real survey. Taking the best fitting value for the quantity
$b_{\rm r} S^{\rm G}_{3} $, and the skewness measured for the
reference sample, we can obtain a best fitting value for the nonlinear
bias, $c_{2}^{'}$, using Eq.~\ref{eq:c2'}. The error on the best
fitting value is obtained by adding the errors on $b_{\rm r} S^{\rm
G}_{3} $ and the skewness of the reference sample in quadrature,
ie. assuming $b_{\rm r} S^{\rm G}_{3} $ and $S^{\rm G*}_{3}$ are
uncorrelated. The resulting values of the nonlinear bias and the error
on the best fit are listed in Table~1.

The results in Table~1 reveal substantial differences in the
hierarchical amplitudes obtained for red and blue galaxies and in the
way in which these amplitudes change with luminosity. Overall, we find
that blue galaxies display smaller hierarchical amplitudes than red
galaxies.  Red galaxies show a significant change in $S_{\rm p}$ with
luminosity, whereas blue galaxies show no such trend.  Assuming
Gaussian errors, we find a 3-$\sigma$ shift in the best fitting $S_3$
between the faintest and brightest samples of red galaxies.  In view
of the relatively small baseline in luminosity over which we can
perform such a fit, a factor of six in median luminosity moving from
the faintest to the brightest volume limited sample, it is remarkable
to see such a clear change in the higher-order clustering. These
results are in good agreement with those found in Croton
et~al. (2004b) (see their Fig.~10). It is also noteworthy that the
relative bias $b_{\rm r}$ quoted in Table~1, shows the opposite
tendency to that displayed by the hierarchical amplitudes: there is
little dependence of the linear relative bias on luminosity for red
galaxies and a clear trend for blue galaxies.

Finally, it is encouraging to note that the conclusions we draw from
our counts-in-cells analysis are consistent with those we reported in
Gazta\~{n}aga et~al. (2005, SAGS-II), who measured the 3-point
correlation function for different triangle shapes and sizes
(i.e. without averaging over the volume of a cell).  In particular
Fig.7 in SAGS-II shows how for both equilateral and elongated
triangles, late galaxies show little luminosity segregation in $Q_3$,
while early galaxies show a strong segregation, with $Q_3$ increasing
with decreasing luminosity (see also Nishimichi et~al. 2007).

\section{Summary}
\label{conclusion}

This paper complements and extends previous clustering analyses of the
2dFGRS, taking advantage of the availability of $R_{\rm F}$-band
photometry to study the spatial distribution of galaxies as a function
of {\em both} their colour and luminosity. We use a counts-in-cells
approach to estimate the count probability distribution function
(CPDF) of red and blue galaxies in different luminosity bins, spanning
a factor of six in luminosity around $L^{*}$. From the CPDF, we
estimate the volume-averaged higher-order correlation functions,
$\bar{\xi}_{\rm p}$, the hierarchical amplitudes, $S_{\rm p}$, and the
reduced void probability function for our galaxy samples, $\chi$. We
use our measurements of the hierarchical amplitudes to constrain the
linear and nonlinear bias parameters in a simple model of galaxy
clustering.

The main results of this paper can be summarised as follows: 

\begin{itemize}

\item {\it Colour segregation in $\bar{\xi}_{\rm p}$:} 
Blue galaxies show significantly lower clustering amplitudes than red
galaxies. This trend holds up to the five point correlation function,
extending the results previously reported for two-point correlations
(e.g. Norberg et~al. 2002a; Madgwick et~al. 2003; Zehavi et~al. 2005).

\item {\it Luminosity segregation in $\bar{\xi}_{\rm p}$:} 
Blue galaxies display a monotonic increase of clustering strength with
luminosity, mirroring the results found for the two-point function of
2dFGRS galaxies (Norberg et~al. 2002a).  The behaviour of the red
population is more complicated. Faint and bright red galaxies are more
clustered than $L^{*}$ red galaxies. A hint of this behaviour was
previously found for the two-point correlation function of early
spectral types in the 2dFGRS (which loosely correspond to the red
sample in this paper) by Norberg et~al. (2002a).

\item {\it Luminosity segregation in $S_{\rm p}$:} 
We fit scale-independent models to the hierarchical amplitudes which
take into account the full covariance between adjacent bins.  From
this we conclude that the hierarchical amplitudes of blue galaxies
show little or no dependence on galaxy luminosity. On the other hand,
the hierarchical amplitudes of red galaxies vary strongly with
luminosity. These new results, split by colour, explain why Croton
et~al. (2004b) found only a weak dependence of the $S_{\rm p}$ on
luminosity.

\item {\it Superstructures in the 2dFGRS.} 
The hierarchical amplitudes for the red galaxies seems to be more
affected by the presence of the superstructures. After removing them,
there is still significant scale dependence of the $S_{\rm p}$, with
opposite gradients found for red and blue galaxies. These results
imply that the flatness of the $L^{*}$~$S_{\rm p}$ measurements of
Croton et~al. (2004b) and Baugh et~al. (2004), after removing the
superstructures, are due to a fortuitous cancellation of the different
trends with scale for red and blue galaxies.

\item {\it The reduced void probability function.} 
Croton et~al. (2004a) found that the reduced void probability function
measured for 2dFGRS samples defined by luminosity displayed a
universal form, and matches the one predicted by a negative binomial
distribution.  Splitting the samples by colour, we find that blue
galaxies show the universal reduced void probability function
consistent with the negative binomial, but red galaxies do not. The
deviation from the negative binomial model is largest for faint red
galaxies. This result is seemingly at odds with previous
interpretations of the success of the negative binomial model, in
which it was suggested that galaxy peculiar motions were the primary
agent behind the form of the reduced void probability function.

\item {\it Linear and non-linear bias.} 
Fry \& Gazta\~{n}aga (1993) introduced a simple model of galaxy bias
in which the density contrast in galaxies is written as a Taylor
expansion of the density contrast in the underlying mass. The first
order bias term in this expansion is the common linear bias and the
second order term is called the quadratic or nonlinear bias. We use
our measurements of the variance and the skewness ($S_{\rm p}$) to
extract the linear and nonlinear bias parameters relative to the
clustering in a reference sample. The reference sample is the one for
which we are able to make our best measurements of galaxy clustering.
In this paper, galaxies with magnitudes in the range $-21< M_{b_{\rm
J}}-5\log_{10}h < -20$ are treated as the reference sample; there is a
reference sample for red galaxies and one for blue galaxies.  The
relative bias parameters extracted for red and blue galaxies are
different. The faintest red galaxies we consider have a linear bias
consistent with that of the reference sample; red $L^{*}$ galaxies
have a linear bias below unity. For blue galaxies, the linear bias
increases with luminosity. The relative non-linear bias for red
galaxies is positive whereas that extracted for blue galaxies is
negative. In all cases, these offsets are significant at the
$1-2\sigma$ level.
\end{itemize}

Our measurements of the higher-order clustering of galaxies as a
function of colour and luminosity provide valuable new constraints on
models of galaxy formation. While the colour of a galaxy is determined
by its star formation history, the clustering of galaxies, at least on
the scales probed in this paper, is driven primarily by the mass of
the host dark matter halo and to a lesser extent by the formation
history of the halo (Gao et~al. 2005; Harker et~al. 2006; Wechsler
et~al. 2006; Croton, Gao \& White 2007).  Semi-analytical models of
galaxy formation make {\it ab initio} predictions for the star
formation histories of galaxies in a cosmological setting (see the
review by Baugh 2006). This is done through simple physical
prescriptions that describe those aspects of galaxy evolution believed
to be important, including the rate of gas cooling, the timescale for
star formation in galactic disks, mergers between galaxies, and
feedback processes such as heating by supernova explosions or the
accretion of material onto supermassive black holes. These processes
are poorly understood and consequently, there is no unique way in
which to model them.

At present there are a number of galaxy formation models which follow
the evolution of disks and spheroids from high redshift to the present
day (e.g. Baugh et~al. 2005; Croton et~al. 2006; Bower et~al. 2006;
Cattaneo et~al. 2006; De Lucia \& Blaizot 2007). The parameters that
constrain the simple physics assumed in these models are typically set
without reference to galaxy clustering. Higher-order clustering and
non-linear bias measures are sensitive to how galaxies populate groups
and clusters, and their periphery.  We therefore look towards such
statistics to further constrain and discriminate between different
possible (and plausible) implementations of the galaxy formation
physics.  The higher-order spatial distribution of galaxies provides
an additional window through which we can hope to understand the
complex physics governing galaxy evolution.

\section*{Acknowledgements}

The 2dFGRS was undertaken using the two-degree field spectrograph on
the Anglo-Australian Telescope. We acknowledge the efforts of all
those responsible for the smooth running of this facility during the
course of the survey, the 2dFGRS team for carrying out the
observations and releasing the data and also the support of the time
allocation committees during the project. This work was supported in
part by NSF grants AST00-71048 and AST05-07428, the Royal Society
through the award of a Joint Project Grant, and by the European
Commission through the ALFA-II programme's funding of the Latin
American European Network for Astrophysics and Cosmology (LENAC). DC
is supported by a DEEP2 postdoctoral fellowship. PN acknowledges
receipt of a PPARC PDRA fellowship held at the IfA.  EG acknowledges
the Spanish Ministerio de Ciencia y Tecnologia (MEC), project
AYA2006-06341 with EC-FEDER funding, research project 2005SGR00728
from Generalitat de Catalunya, the Galileo Galilei Institute for
Theoretical Physics for hospitality, and the INFN for partial support
during the completion of this work. CMB is funded by a Royal Society
University Research Fellowship.

\bibliographystyle{mnras}

\end{document}